\documentclass[sigconf]{acmart}
\author{Lenore M. R. Mullin}
\email{lmullin@albany.edu}
\affiliation{
 \institution{College of Engineering and Applied Sciences\\ University at Albany, SUNY }
  \streetaddress{1400 Washington Avenue}
  \city{Albany}
  \state{NY 12222}
  \country{USA}
  \postcode{12222}
}

\AtBeginDocument{%
  \providecommand\BibTeX{{%
    \normalfont B\kern-0.5em{\scshape i\kern-0.25em b}\kern-0.8em\TeX}}}

\usepackage{listings}
\usepackage{color}
\usepackage{url}
\definecolor{dkgreen}{rgb}{0,0.6,0}
\definecolor{gray}{rgb}{0.5,0.5,0.5}
\definecolor{mauve}{rgb}{0.58,0,0.82}

\acmConference[HLPP2023]{}{June 29-30, 2023}{Cluj-Napoca, Romania}

\begin{document}
\pagenumbering{arabic}
\title{From array algebra to energy efficiency on GPUs}

\subtitle{Data- and hardware shapes with dimension-lifting to optimize memory-processor layouts.}

\begin{abstract}
We present a new formulation for parallel matrix multiplication (MM) to 
out-perform the standard row-column code design.  This algorithm is formulated in the MoA formalism (A Mathematics of Arrays, \cite{mul88}\cite{hains1993}) and combines an array view of hardware 
({\em dimension-lifting} to extend indexing to physical memory/processing units), with a contiguous data layout derived from static transformations. 
This view of a hardware-software model is thus a {\em bridging model} in the sense of Valiant's BSP. 
OpenACC code was derived from the MoA expression's normal form, producing optimal block sizes 
using the static information of types and shapes. 
Experiments were run on Nvidia V100 GPUs and reveal energy consumption which is quadratic in N, i.e. linear 
in the size of matrix. 
More generally this approach may be an ideal way of formulating, optimizing, and mapping array algorithms to embedded hardware. This work builds upon recently published results of NREL scientists.
 


\end{abstract}
\maketitle









\section{Introduction}
Embedded Systems, up until recently, were  referred to as Embedded Digital Systems typically used in Signal Processing. Optimizations  came from advanced HDL compilers\cite{AshendenPeterJ2008DGtV}. Algorithms were well defined and only the best engineers knew how to exploit the tools to optimize the special purpose hardware they targeted.  Today, just about every device in our homes, our cars, and at work, rely on reliable, low cost, energy-efficient, embedded systems.  Consequently, software programmers are seeking high level languages and tools to formally describe, and ideally verify, algorithms in their domains while compiling to FPGAs, without requiring an extensive engineering background. Current research addresses this problem\cite{ACM0}, however things like "loop optimizations" are done after the specification of algorithms in a high level language like Python or C and use  Pragmas. Certainly all these initiatives should continue using the theories and methodologies 
available, i.e., incremental changes to existing theories and methods\cite{6844463,DBLP:journals/corr/abs-2102-07952}. But, as suggested recently in CACM\cite{atthetop}, there is room at the top for new ideas, and theories. 
We propose MoA (A Mathematics of Arrays \cite{mul88}\cite{hains1993}) to play this role of 
a formalism to bridge high-level functional algorithm descriptions with hardware and memory shapes 
and sizes. 
Recent publications\cite{Mm1,Mm2} show that MoA's formulation of Matrix Multiplication(MM), out-performs direct application of the Linear Algebra definition: a row of A with column of B to obtain a component in C. Given MM is at the heart of most domains, further studies are justified to explore how contiguity leads to optimizations. MoA-derived parallel code 
for MM has been measured to outperform existing libraries. 
In this paper we explain how this result is obtained and how 
it is based on a general methodology. 


This work in progress research  builds upon recently published work by National Lab (NREL) scientists\cite{Mm1,Mm2}  Experiments were run on Nvidia V100s while apriori theorizing about optimal block size using  shapes and types of arguments. MoA conjectured a uniform view over 20 years ago at a High Performance Embedded Computing Conference (HPEC) conference\cite{hpec,radar}. The MoA MM has also been 
efficiently implemented in an FPGA\cite{info13110528,DBLP:journals/information/GroutM22}.

    

\section{MoA Matrix Multiplication}
\label{sec:moamm}

The general matrix-matrix multiplication (\texttt{GEMM}) in MoA  is a special case of the \textit{inner product} for 2-D arrays (matrices), emphasizing that in the MoA MM ALL arrays are accessed contiguously. Define $\mathbf{A}$ as an $m \times n$ matrix, $\mathbf{B}$ as $n \times p$, and $\mathbf{C}$ as $m \times p$. In MoA notation, the shapes of $\mathbf{A, B, \mbox{ and } C} $ are respectively:
\begin{eqnarray}\label{eq:shapes}
    \rho \mathbf{A} &= \langle m, \, n \rangle \nonumber\\
    \rho \mathbf{B} &= \langle n, \, p \rangle \\
    \rho \mathbf{C} &= \langle m, \, p \rangle \nonumber
\end{eqnarray}
so the valid indices of the matrices that are bounded by shapes:
\begin{equation}\label{eq:indices}
\forall \; i, j, k \ni \begin{cases}
    0 \leq i < m \\
    0 \leq j < p \\
    0 \leq k < n
\end{cases}     
\end{equation}
The MoA Operational Normal Form (ONF) for \texttt{GEMM}\cite{Mm1,Mm2} is given by the following expression:
\begin{eqnarray}\label{eq:dnf}
{\bf C}[(i \times p) + j] := \sum_{k=0}^{n-1} {\bf A}[(i\times n)+k] \times {\bf B}[(k\times p)+ j]
\end{eqnarray}
This is a "generic row-major form" whose meaning is that $\forall i:(0..m-1)$, $ \forall j:(0..p-1)$, $ \forall k:(0..n-1)$ 
the content of memory from the initial address {\bf @C} of array C is:
\begin{eqnarray}
{\bf @C}+(i \times p) + j := \sum_{k=0}^{n-1}({\bf @A}+(i\times n)+k) \times ({\bf @B}+(k\times p)+ j)
\end{eqnarray}
where {\bf @A} (resp. {\bf @B}) is the initial address of array A (resp. B). 
Let the following notation denote the 2-d Matrix Multiplication defined by MoA's \textbf{inner product} definition in (\ref{eq:dnf}),
\begin{equation}\label{eq:ip}
    \mathbf{C} = \mathbf{A} \bullet \mathbf{B}
\end{equation}
In other words the theory's inner product operator on 2D arrays is equivalent to the definition of 
matrix multiplication. The appendix to this paper summarizes core elements of MoA and details why the MoA MM is different in terms of memory access patterns as compared to the classical definition, a row of A with a column of B.  

\begin{figure}
    \includegraphics[width=3in]
    {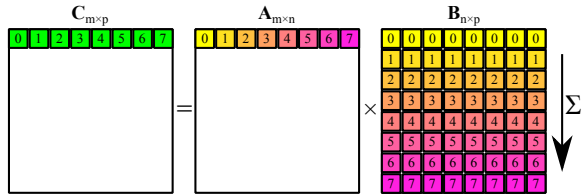}
    \caption{(Color online) Inherent parallelism of the MoA matrix-matrix multiplication (\texttt{GEMM}). The different colors and numbers on matrices $\mathbf{A}$ and $\mathbf{B}$ represent the pairings for the $k$ index involved in the scalar-vector multiplication, e.g. the matrix element labeled 0 in $\mathbf{A}$ is multiplied with the row vector labeled 0 in $\mathbf{B}$. A sum reduction is then performed over the resulting row vectors to yield the $i$th row in $\mathbf{C}$.}
    \label{fig:mm}
\end{figure}

Equation \eqref{eq:dnf} is the generic code for a sequential program in MoA. Figure \ref{fig:mm} illustrates the inherent \textbf{parallelism} of the MoA \texttt{GEMM} algorithm. In each $i$th row of the resultant array $\mathbf{C}$, each scalar-vector operation involving the column index $j$ is independent of each other. The $i$th row of $\mathbf{C}$ is contiguously filled in by the summation of scalar-vector multiplications involving each matrix element at the $i$th row and $k$th column  of $\mathbf{A}$ (the scalar) with each $k$th row of $\mathbf{B}$ (the vector) obtained by accessing the arguments contiguously. A row-wise sum reduction is then applied over the $k$ index to yield the final answer as the $i$th row in $\mathbf{C}$.

\begin{figure}
    \includegraphics[width=3in]
    {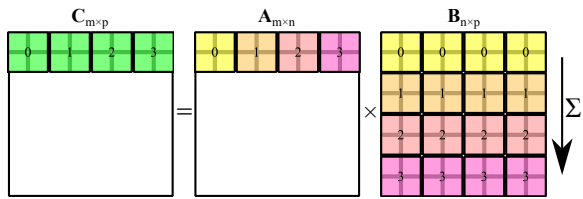}
    \caption{(Color online) Inherent parallelism of the blocked MoA general matrix-matrix multiplication (\texttt{GEMM}) algorithm. Just like the scalar \texttt{GEMM} design (Fig. \ref{fig:mm}, the block-block multiplications between different blocks in the row of matrix $\mathbf{A}$ and the corresponding rows of blocks in matrix $\mathbf{B}$ are independent of one another, and a sum reduction over the rows of blocks are performed to yield the final answer of blocks in matrix $\mathbf{C}$.}
    \label{fig:omega}
\end{figure}

Figure \ref{fig:omega} shows how the {\em blocking} algorithm applies the inner product, Equation \eqref{eq:ip},  to blocks in a round robin, row-major order, just like the scalar version in Figure \ref{fig:mm}, but this time summing blocks (subarrays) of partial sums. With each matrix block just large enough to fit in the L1 cache, each block operation is performed contiguously, round robin style, and efficiently. 

The original feature of our algorithm is that 
\begin{enumerate}
    \item it was symbolically derived from a functional specification and MoA equivalence laws,  
    \item a normal form exists and can be derived by confluent rewriting, much like parallel functional or skeleton-based programs, 
    \item its resulting normal form can be expressed as a C code whose structure is a 
    parallel execution of loop-nests that expresses the processor layout and memory-block accesses. 
\end{enumerate}
In the next section we summarize this program derivation and the experimental performance analysis.

\section{From Normal Form to C Program to Dimension Lifting}
First, the generic design in Equation (\ref{eq:dnf}) is implemented in a C program illustrated in Figure  \ref{ip.c}.
\begin{figure}[htbp]
\begin{lstlisting}
void ip(double *C, double *A, double *B,
  int sizel, int sizer, int sizeres, int np, int shr0)
 {int i,j,sigma;
  for (i=0;i<sizel;i++)
    {for (sigma=0; sigma<shr0; sigma++)
      {for (j=0;j<sizer;j++)
	    {
          C[j+i*sizer]=C[j+i*sizer]
                      +A[(i*shr0)+sigma]*B[(sigma * sizer)+j];
	    }}}}
 \end{lstlisting}
 \caption{From Generic Form to C Program (ip.c)}
 \label{ip.c}
 \end{figure}

\subsection{Dimension Lifting to Machine Coordinates}
\begin{definition}
Dimension lifting is defined by systematically partitioning each shape component into 2, thus lifting the dimension of the problem as each partitioned shape is used to identify an architectural resource, like processors.
\end{definition}
Dimension lifting is a way to abstract and unify an algorithm with the architecture it maps to. Through a systematic unified view guided by MoA, it become possible to fully automate algorithm optimizations through Psi Reduction (index composition), done prior to compilation. Then through dimension lifting, all parallelism is revealed s.t. costs and optimizations become possible\cite{DBLP:journals/corr/abs-0811-2535,pythonmoa,hpec,wilmul}. The C programs presented herein are augmented with OpenACC, noting that after parallelism is revealed, pragmas can be easily added with confidence of competitive performance with CUDA\cite{wilmul}. 

Dimension lifting over the rows of A and C, the i loop, reveals parallelism and assigns an index
to processors. This was done  as illustrated in Figure \ref{ip_rows.c}. 

\begin{figure}[htbp]
\begin{lstlisting}
   void ip_rows(double *C, double *A, double *B,
	 int sizel, int sizer, int sizeres, int np, int shr0)
 {int i,j,k,ip,sigma;
   // for (i=0;i<sizel;i++)
   for (k=0;k<np;k++)
    {for (ip=0;ip<(sizel/np);ip++)
     {for (sigma=0; sigma<shr0; sigma++)
      {for (j=0;j<sizer;j++)
	    {
    // C[j+i*sizer]=C[j+i*sizer]
    //               +A[(i*shr0)+sigma]*B[(sigma * sizer)+j];
	C[j+(ip+(sizel/np)*k)*sizer]=C[j+(ip+(sizel/np)*k)*sizer]
    +A[((ip+((sizel/np)*k))*shr0)+sigma]
      *B[(sigma * sizer)+j];
	    }}}}}
\end{lstlisting}

\caption{Dimension Lifting over the Rows of A (ip\_rows.c)}
\label{ip_rows.c}
\end{figure}

Dimension lifting over the columns of B reveals parallelism also. This would break up the j loop as illustrated in Figure \ref{ip_cols.c}. Mapping each row of B could be in groups of 8 e.g. to a vector register or a group of threads. 
\begin{figure}[htbp]
\begin{lstlisting}
    void ip_cols(double *C, double *A, double *B,
	 int sizel, int sizer, int sizeres, int np, int shr0,int rsize)
 {int i,j,jp,kp,sigma;
  for (i=0;i<sizel;i++)
   {for (sigma=0; sigma<shr0; sigma++)
	{for (jp=0;jp<(sizer/rsize);jp++)
	  {for (kp=0;kp<rsize;kp++)
	//{for (j=0;j<sizer;j++)
	{
	//  C[j+i*sizer]=C[j+i*sizer]+A[(i*shr0)+sigma]*B[(sigma * sizer)+j];
  C[((jp*rsize)+kp)+i*sizer]=C[((jp*rsize)+kp)+i*sizer]
  +A[(i*shr0)+sigma]*B[(sigma * sizer)+((jp*rsize)+kp)];
	    }
    //}
	}}}}
\end{lstlisting}
\caption{Dimension Lifting over the Columns of B (ip\_cols.c)}
\label{ip_cols.c}
\end{figure}
Finally, the sigma loop is broken up creating the block. This 
necessitates another addition loop to add up the blocks realizing
Figure \ref{fig:omega}.
\subsection{Hardware and Software}
\begin{table}[h]
    \centering
    \caption{Relevant numbers for the memory and execution unit hierarchy for the NVIDIA V100. }
    \label{tab:gpu}
    \begin{tabular}{|l|c|c|}
        \hline
        \multicolumn{2}{|l|}{} & V100  \\
        \hline
        L1 cache size & $x$ (KiB) & 128  \\
        L2 cache size & $y$ (MiB) & 6  \\
        Global memory size & $z$ (GiB) & 16 or 32  \\
        \hline
        Number of SMs & $N_\mathrm{SM}$ & 80  \\
        \hline
        \end{tabular}
\end{table}
\subsection{Conjectures about performance}
Thinking of the parallel execution of ip\_rows.c we find that 


\begin{enumerate}
    \item from an MoA, shape point of view, the L1 cache size will dominate the block sizes of the A, B, and C matrices performing summations of scalar-vector products; the block sizes being the total number of components in the block, that is, the product of the shape vectors of each matrix block ( square or non-square), and
    \item the block size should change when shared memory is used between SM's and their L1s.
\end{enumerate}


\subsection{Technical Details} 
In this section we describe the derivation 
of parallel (OpenACC) code from 
the C-like code that we consider to be the 
machine-independent operational semantics of 
our algorithm. It applies the data-parallel 
principle of mapping data to machine elements 
and making references/communications dependent 
on that mapping. To simplify the implicit 
cost- and machine abstractions we retain only 
the most critical factor in mapping the 
algorithm to the GPU architecture, namely the 
exact block sizes {\em sizel, sizer} that 
dimension lifting has created symbolically. 
The optimization objective is 
to make those values as close as possible to 
the GPU cache sizes. This approach to cost 
optimization is well-suited to computation on 
a single GPU as our results show. 

Other 
cost-models could also be applied to the 
MoA-derived operational semantics, for example 
logP \cite{culler1993} (for multi-core 
implementations), BSP \cite{mccoll1999} 
or its heterogeneous variants \cite{williams2000} that account for communication costs. The operational semantics 
of our C-like code and cost-modeling are 
similar in spirit to the resource-aware 
single-assignment C of \cite{grelck2020}.  

In the current experiment, all memories, both local and global with speeds and sizes,  
are thus 
formulated relative to shapes of arrays and their blocks. 
\noindent
The V100 has 80 Streaming Multiprocessors (SMs) or $80 \times 32$KB.
Each SM has 32KB:
$32 \times 2^{10}$ or
$2^2 \times 2^3 \times 2^{10}$ Bytes or $2^{15}$ bytes. Consequently, the block size chosen, i.e. rows and column size of doubles should be this amount of bytes. Although initially blocks are square, it is the number of components in the block that matter. 
Based on  experiments, the best performance is when each block is 32 by 32 doubles (or could have been 16 by 64, 8 by 128,  etc.). That is $2^2 \times 2^3 \times 2^2 \times 2^3 \times 2^3$. This is, $2^{10}\times 8$ or $1024 \times 8$, or 8KB . There are three blocks per SM: for A, B, and C, so  24KB. The block size must be less than or equal to 1 L1.

The next best block size was 64 by 64 doubles, so each block size has another power of 2. So,
$2^{10} \times 8 \times 4 = 32$KB times 3 means 3 SMs must be in use, knowing the L1 with shared memory is 128KB. 

Why does the block size change from 32 by 32 to 64 by 64? Perhaps investigating
how  the L2 cache is divided amongst 80 SMs may bring insights.  The V100 has a 6MB L2 or $6 \times 10^{20}$Bytes.
Assume as the matrix size exceeds the size of L2, that L2 size blocks would be loaded from global memory, this is another "dimension lifting" of the overall algorithm. There is a change in block size when the matrix size becomes around $9 \times 1024$ by $9 \times 1024 \times 2^3 \times 3$, 
i.e. 3 matrices with doubles. 
This is approximately, $81 \times 3 \times 2^{20} \times 2^3$ or approximately $8 \times 32 \times 2^{20} \times 2^3$ or $2^8 \times 2^{20} \times 2^3$ which is about $2 \times 2^{30}$ or 2 GB. In this case, what might be the case for the V100 with 8 GPUs per node and 16 GB of global memory, is that each GPU is allocated 2GB. For the max size of 6GB, the V100s may know how to use memories from the other GPUs. 


The maximum matrix size in the experiments summarized by our plots is 3 16000 by 16000 doubles or $2^4 \times 2^{10} \times 2^4 \times 2^{10} \times 3 \times 2^3$. This is $2 \times 2^{30} \times 3$ or $6\times 2^{30}$ or 6 GB. 
What is definitely known is that the block size is dependent only on the L1 size and it increases by a factor of two at the 9K by 9K interval and so conjectured to use multiple SMs with shared memory.

The above calculations are not only a trace of this systematic 
code derivation but an attempt to identify all variables needed to formulate the desired abstractions algebraically. 
Future work will investigate the combination of automated 
MoA normal calculation with symbolic (hence portable) 
cost optimizations. 

\subsection{Performance plots} 
We measured speed and energy consumption as a function of (one of two) architectures, block sizes and 
matrix sizes. 


The best time was achieved with a 32KB by 32KB block size changing to 64KB by 64KB. Best time block size was correlated to best Energy block size. There is an inverse correlation with Power and Heat. That is, the worst Time and Energy block sizes occurred when Power and Heat were the best.

\newcommand\w{.36}
\begin{figure}[p]
 \centering
  \includegraphics[width=\w\textwidth]{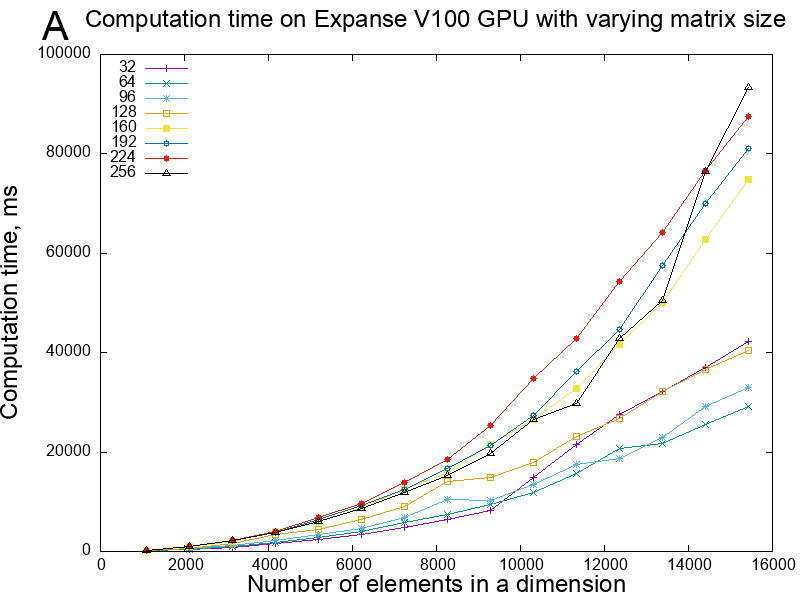}
 \hspace{.1cm}
  \includegraphics[width=\w\textwidth]{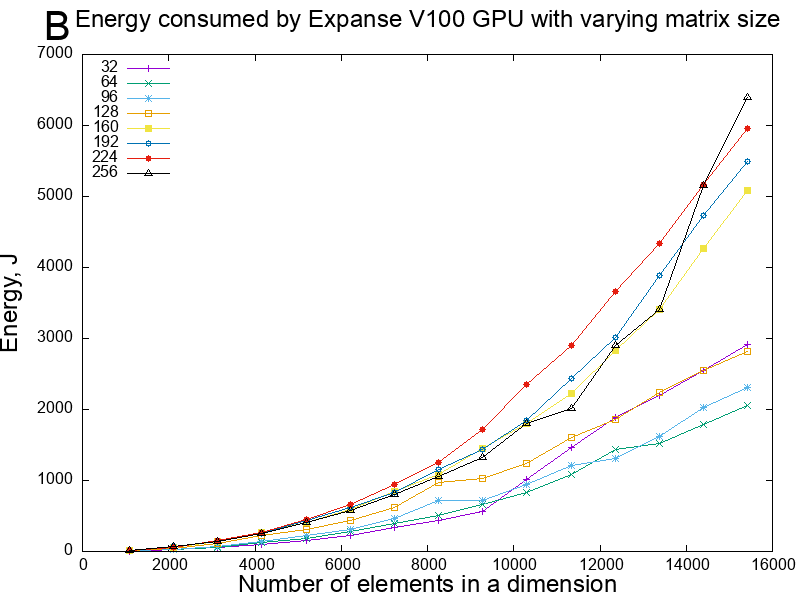}
\caption{Time and Energy: V100 with 32GB RAM on Expanse.}
\end{figure}

\begin{figure}[p]
\includegraphics[width=\w\textwidth]{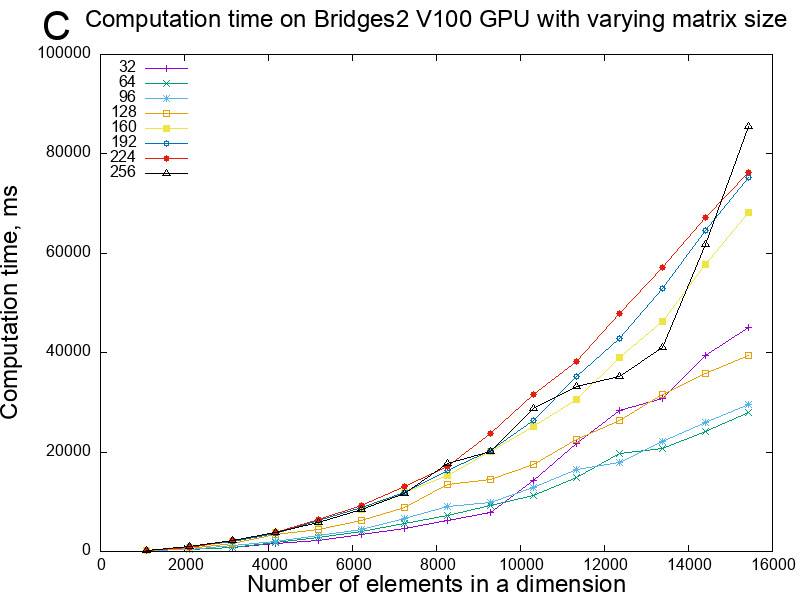}
  \hspace{0.1cm}
\includegraphics[width=\w\textwidth]{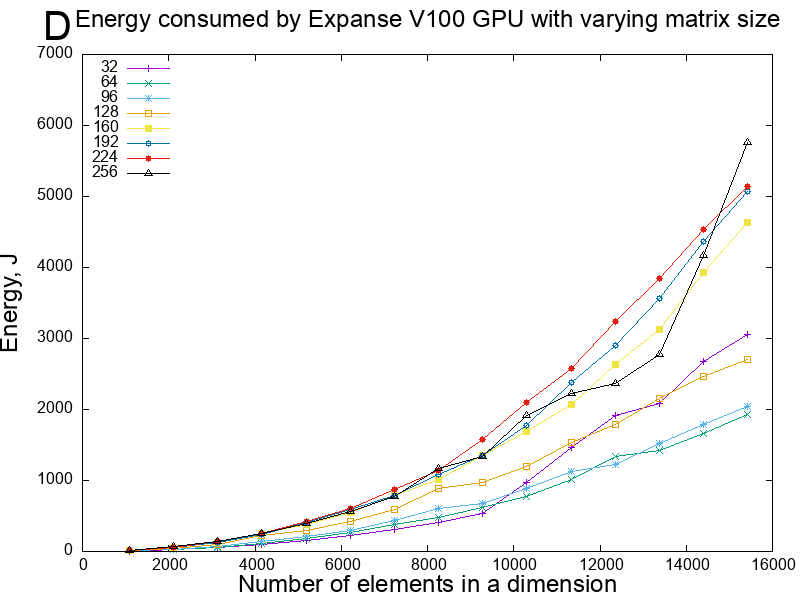}
\caption{Time and Energy: V100 with 16GB RAM on Bridges1.}
\end{figure}

\begin{figure}[p]
\includegraphics[width=\w\textwidth]{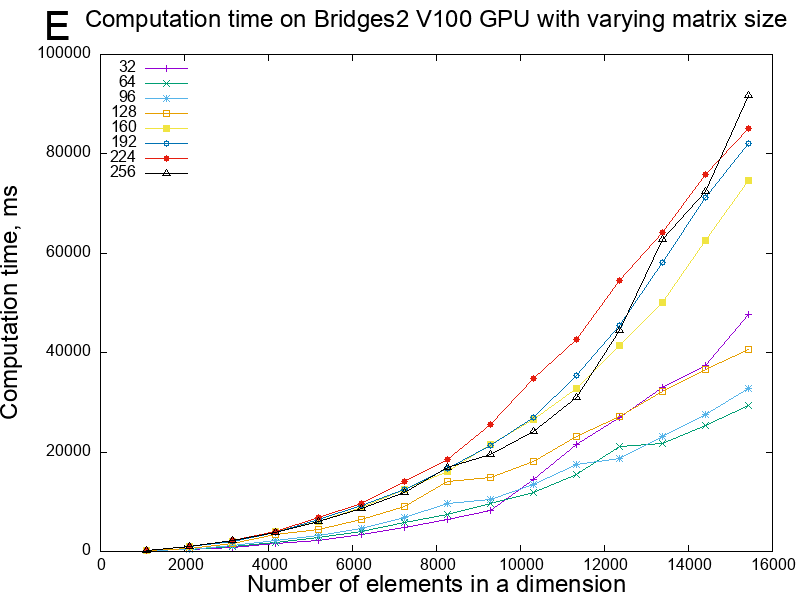}
  \hspace{0.1cm}
\includegraphics[width=\w\textwidth]{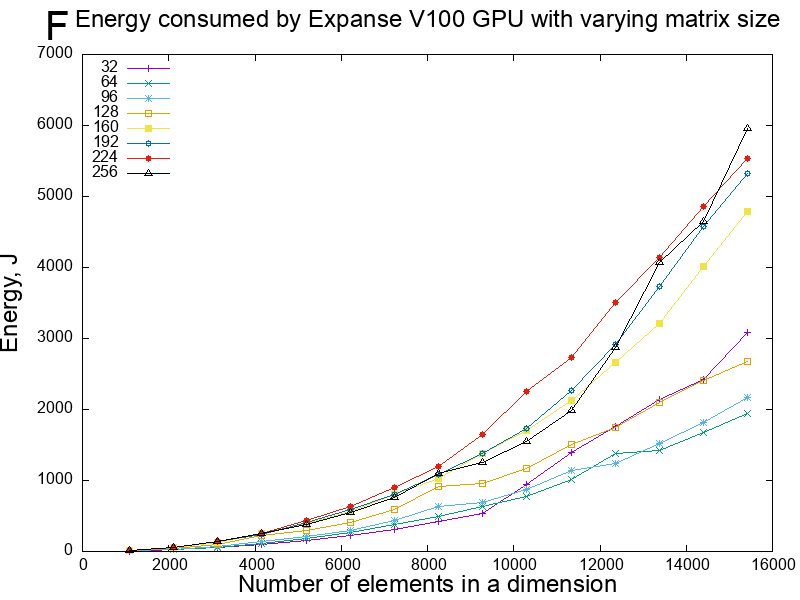}
\caption{Time and Energy: V100 with 32GB RAM on Bridges2.}
\end{figure}


\begin{figure}[p]
  \centering
  \includegraphics[width=\w\textwidth]{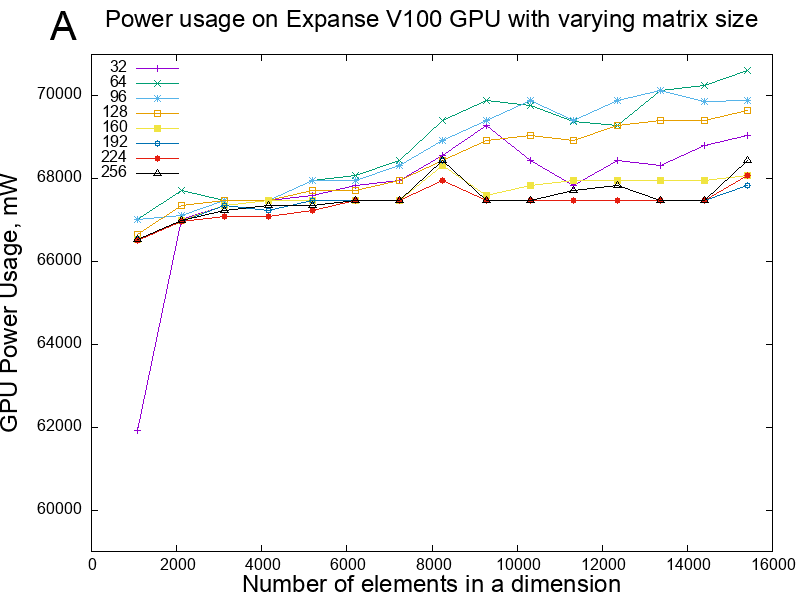}
  \hspace{0.1cm}
  \includegraphics[width=\w\textwidth]{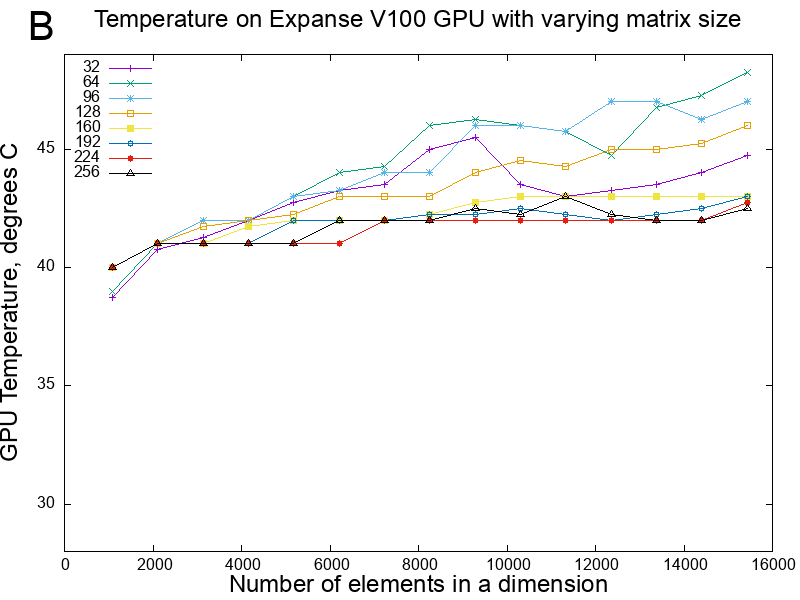}
  \caption{Power and Temperature: V100 with 32GB RAM on Expanse.}
\end{figure}

\begin{figure}[p]
  \includegraphics[width=\w\textwidth]{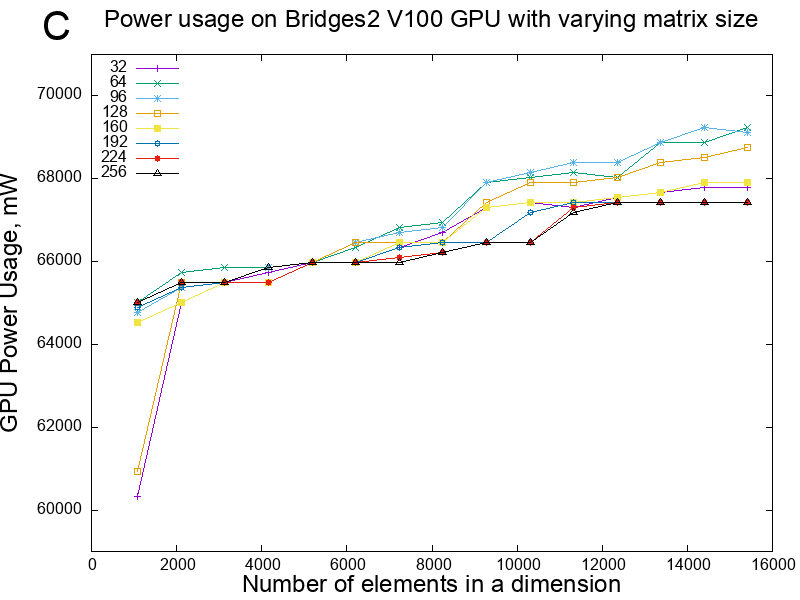}
  \hspace{0.1cm}
  \includegraphics[width=\w\textwidth]{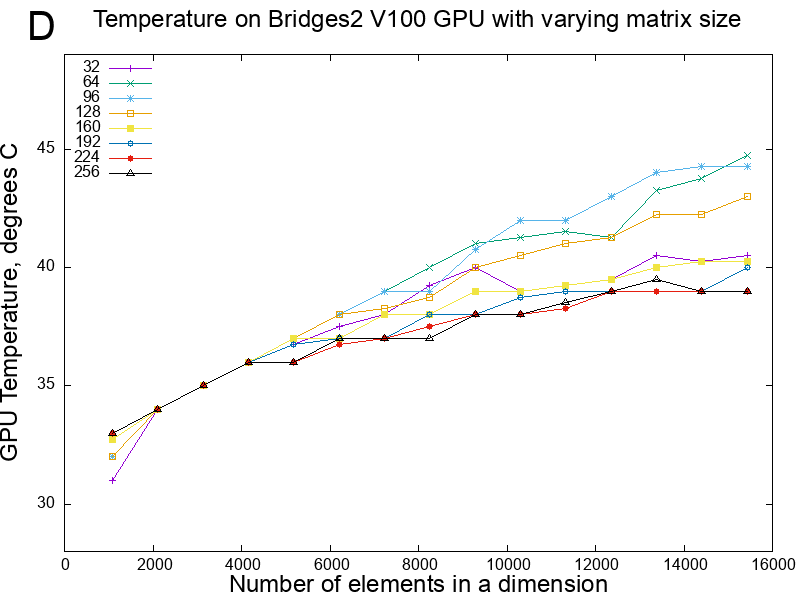}
  \caption{Power and Temperature: V100 with 16GB RAM on Bridges1.}
\end{figure}

\begin{figure}[p]
  \includegraphics[width=\w\textwidth]{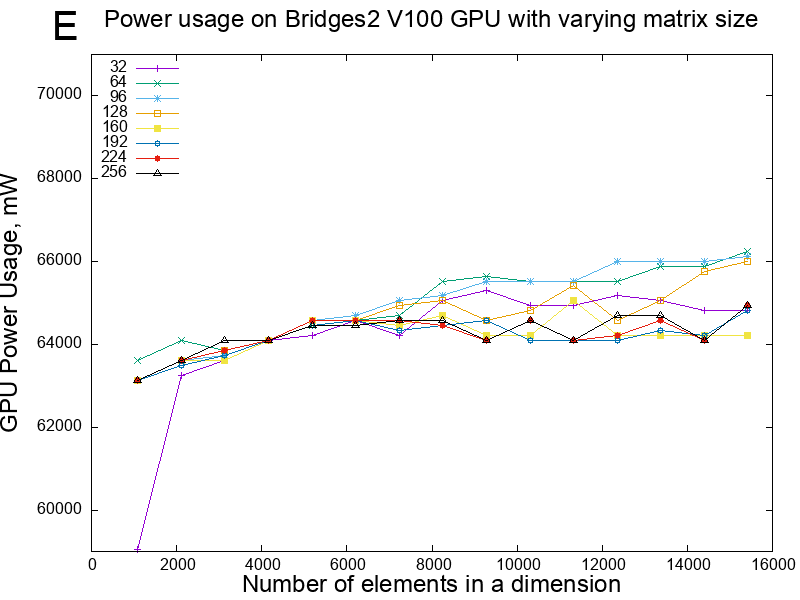}
  \hspace{0.1cm}
  \includegraphics[width=\w\textwidth]{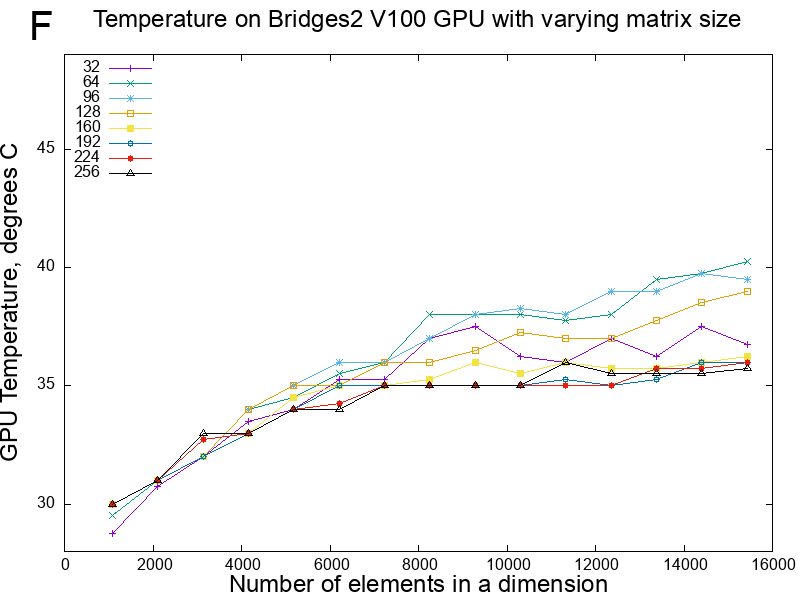}
  \caption{Power and Temperature: V100 with 32GB RAM on Bridges2.}
\end{figure}


%
%
Outcome is preliminary but optimistic. Initial conjectures were validated. 
\subsection{Observations}
\subsubsection{Performance: Time and Energy}
Block size is defined by the L1 memory. Optimal L2 and feeding the L2 is hierarchically related to the Number of SMs and size of their L1s with shared memory between them. Maximum matrix sizes per GPU are related to the number of GPUs and the Global Memory. 
There is a direct correlation between optimal Energy, optimal Time, and block size. 

\subsubsection{Performance: Power and Heat}
On all machines, Power and Heat had an inverse relationship with Time and Energy. 
This implies  knowledge of the L1 cache size, the number of SMs, amount of shared memory, the size of the L2 and global memory, all deterministic.
\subsubsection{Performance: Power and Time}
Notice that even though energy consumption and running time plots seem to be exactly the same, power doesn't vary nearly as much as time results do. When calculating energy the difference in power usage is much less noticeable compared to huge variations in time, taking the data from the  Expanse 32 block size experiment as an example. 

The ratio between maximum and minimum value for power usage is 69,038.2 / 61,933.3 = 1.115. This is by far the largest ratio compared to other block sizes, because the first datapoint for the power usage chart for 32 block size is a clear outlier. The ratio between maximum and minimum value for time is 42,313.3 / 111.8 = 378.3. So while power usage increases by ~10\%, time results increase by 378 times. If we calculate the same ratio for energy data we get 421.7, so power usage does contribute, however looking at the graph this contribution is not noticeable at all and all the lines on the energy plot look just like the ones on the power usage plot.
\subsection{Collecting the Independent Variables}
With a goal of using MoA to abstract the algorithm, the architecture, including all communications, shapes of the algorithm need to be "dimension lifted", i.e. partitioned to reflect the architectural components with communications desired, so that each has an index. This research requires more experimental confirmations of our theoretical predictions but constitutes a clear proof 
of concept for systematic implementation of an algorithm using
algebraic normal form combined with a block-size to GPU-caches cost model. 
\subsection{Outperforms Standard Libraries?}
This paper begins with a claim that it "outperforms" other algorithms. 
But the experiments described here only measure variants of the same (our) algorithm. So why is that claim realistic? 
\begin{itemize}
  \item We view and analyze all data and resources as contiguous 
no matter what the array layout (row major, column major, sparse, ...) 
and in so doing, maximize performance. The previous control experiments using CPUs and Threads\cite{Mm1,Mm2}, validated  the conjectures of performance improvements. However, performance is not the only criterion for optimality 
of most applications. 

\item 
Maximizing parallelism theoretically minimizes time but is often not realistic 
because of space, processing power and hence energy demand. 
Our Mm algorithm is formulated in terms of the Kronecker product 
so as to combine possibly disjoint operations into a single sequence 
of array operators \cite{Grout18,Grout19}. 
Because of this monolithic design it could be called from a sequential 
host language the way Python is used for combining tensor operators \cite{info13110528}. 
The algorithm's structure can be adapted to support Hadamard product and 
sparse MM. This is discussed in the appendix. 
\end{itemize}

\section{Conclusion}

A new matrix multiplication algorithm was presented with blocking described in terms of {\em 
dimension lifting}, a term coined for the MoA (mathematics of arrays) theory, that formalizes splitting of indices to give an index to an architectural component, thus increasing the dimension of the algorithm. Dimensions increase because MoA views the algorithm {\em and} architectural resources in a uniform Cartesian way. Consequently, the abstraction of a 2-dimensional algorithm, already optimized through MoA's Psi-reduction could be 3-dimensional (e.g. adding processors), 4-d (e.g. adding processors and vector registers), or n-d (whatever resources chosen). This approach internalizes the target hardware shape in the array formalism and 
allows formal, safe and high-level transformations to optimize generated parallel code. 

Reproducible results imply that MoA's contiguous view of memory accesses 
outperforms the best efforts to optimize the classical design\cite{Mm1,Mm2}.

Paper \cite{atthetop,ACM0} has called for new concepts in parallel computing 
with the slogan "there is room at the top". We believe that algebra and 
cost models can provide such high-level concepts to support present and 
future high-performance computing. 

This research will continue to explore how shapes can be used to  predict performance of MoA MM on more GPUs.

\begin{acks}
This work used Nvidia V100s at XPANSE through allocation Start-UP Grant
CIS210035 from the Extreme Science and Engineering Discovery Environment (XSEDE), which was supported by National Science Foundation grant number \#1548562. Special Thanks go to Dima Shyshlov, XSEDE Campus Scholar, for his help in organizing and running experiments and plots and to Mohamed Zahran for his expert advice on GPUs.
\end{acks}

%

\bibliographystyle{ACM-Reference-Format}
\bibliography{conf_paper}

\newpage 

\section*{Appendix: MoA, functional combinators for arrays}
\footnote{The reader preferring an overview is encouraged to just read  the italicized  sentences.}

Linear- and multilinear algebra, matrix operations, decompositions, and transforms are at the core of IoT (Internet of Things), AI (Artificial Intelligence), ML (Machine Learning), and Signal Processing applications. In particular, General Matrix Matrix Multiplication (GEMM) \cite{Mm1,Mm2} and the Kronecker Product (KP) are the most commonly used \cite{kron1} products. The Khatri-Rao (KR), i.e., parallel KPs, was reported to consume 90\% of all IoT computation at Siemens in Munich \cite{Acar16}. 
In AI, Recurrent Neural Networks (RNNs) can be difficult to deploy on resource constrained devices. 
But KPs can compress $\mbox{RNN}$ layers by 16 - 38  times with minimal accuracy loss \cite{Rnn0}. 
Both, MM and KP occur {\bf often and in multiples} \cite{Mullin14} within today's software AI tools, e.g., Tensorflow  and Pytorch, written in Python and NumPy. Due to the use of interpretive languages that are utilized to formulate these tools, even with NumPy, and other software accelerators to speed up interpretation, special purpose hardware is being explored, e.g., GPUs and TPUs, as performance accelerators (co-processing).
Even with all this technology, the necessary speeds and sizes of matrix operations needed are not being realized.

We summarize here the Mathematics of Arrays (MoA) formalism that we use to 
define and transform matrix/tensor algorithms in a declarative way. 
MoA is also used to map operations to parallel hardware by so-called 
dimensional lifting which can express block structures to match the 
hardware in quantity and size of memory units. Computational dependencies 
are explicit but communications are implicit, making MoA an excellent   
basis for high-level parallel programming. 

The theoretical definitions begin with shapes and the Psi indexing function. Together they can define arbitrary array operations that, when composed, yield a semantic/Denotational Normal Form (DNF), which expresses the least amount of computation and memory access, needed for the algorithm while revealing all of its parallelism. 
From the DNF, loops bounds are partitioned, "dimension-lifted", to map to the hardware. MoA views the hardware as arrays: indices of data to indices of machine attributes (registers, memories, processors), and can help minimize costs 
by maximizing locality.
MoA has the Church-Rosser (CR) property \cite{Chetioui19}, and when combined with the Lambda Calculus \cite{Berkling90},  two array programs can be proven equivalent.

\vspace{.2cm}
\noindent
{\em Existing array theories and compiler optimizations on array loops, are proper subsets of MoA. All of NumPy's array and tensor operations can be formulated in MoA. In MoA a single algorithm, thus, one circuit, unifies the Hadamard Product, Matrix Product, Kronecker Product, and Reductions(Contractions) versus four. 
Designs based on this definition can consequently 
consume less circuitry, power, and/or energy. 
This kind of operator unification is well known 
to researchers who use design-patterns, algorithmic skeletons or 
generalized homomorphisms  \cite{legaux2013} to 
express parallel algorithms. But MoA is specific 
for its homogeneous use of the array data type. 
Merging it with the above methodologies could lead 
to block-recursive Strassen-like schemes but that 
is out of scope for this paper. 
}
\subsection*{Shapes and the Psi Function}
Although MoA's algebra was influenced by Iverson's APL language \cite{Iverson62} and Abram's shapes and indexing tables \cite{Abrams70}, it distinguishes 
itself as an equational theory (not only an operational semantics) and 
is built from only two primitives, namely the Psi Function, $\psi$, and 
array shapes. Algebraic structures (group, ring, field, etc.) can be 
added to MoA's indexing calculus. The idea of MoA and Psi Calculus,  starts with a scalar, $\sigma$, and it's shape $\rho \sigma$, an empty vector, denoted by $<>$ or $\Theta$. Since a vector is an array,  it has a  shape, $(\rho <>) \equiv <0>$, the one element vector containing the integer 0, denoting a scalar is a 0-dimensional array. Thus, from the shape, we can determine dimensionality, 
$(\rho (\rho \sigma)) [0] \equiv  <0>[0] \equiv 0$ and the total number of components, $\pi (\rho \sigma)\equiv \pi <> \equiv 1$, in the array. Algorithms on shapes describe how to index arrays with Psi and are defined, such that, Psi takes a valid index vector, or an array
of valid index vectors, and an array as arguments.  For example, in the scalar, $\sigma$, case, $<> \psi \;\sigma$.
Next, 2 scalars(0-d) and operations between them can be considered, with an extension to operations with n-dimensional (n-d) arguments. 

\vspace{.2cm}
\noindent
{\em Scalar operations are at the heart of computation, $\sigma_l f \sigma_r$, and in general for n-d arrays, $\xi_l f \xi_r$, where $f$ is an arbitrary scalar function.}

\vspace{.2cm}
Thus, in general, 

\begin{eqnarray}
 \forall\; \vec i \ni 0 \leq^* \vec i <^* \rho \xi_l \nonumber \\
&& \vec i \mbox{ is a valid index vector.}\nonumber \\
 \vec i \psi (\xi_l f \xi_r) & \equiv & (\vec i \psi \xi_l)\; f\; (\vec i \psi \xi_r) \nonumber 
 \end{eqnarray}
 
 \vspace{.2cm}
\noindent
{\em That is, indexing distributes over scalar operations, or in compiler optimization terms, loop fusion.}
 
\noindent
For a scalar there is only one valid index , $\vec i$  is  $<>$,  the empty vector.
 \[<> \psi  (\sigma_l f \sigma_r)  \equiv  (<> \psi \sigma_l)\; f\;  (<> \psi \sigma_r ) \mbox{(DNF)} \]
 \noindent
 Indexing distributes over scalar operations. \\
 \[\equiv  (\mbox{rav} \;\sigma_l) [\gamma_{row}(<>;<>)] \;f\; (\mbox{rav}\; \sigma_r)[\gamma_{row}(<>;<>)] \]
 \noindent
Relating the DNF to the Operational Normal Form (ONF). 
rav flattens the array in row major.
$\gamma_{row} $ gets the offset using index and shape. \\
\[\equiv  (\mbox{rav} \;\sigma_l)[0] \;f\;(\mbox{rav}\; \sigma_r)[0] 
\mbox{ This is the ONF. }\]
\noindent
Applying the definition of $ \gamma_{row}$  in order to get 0.
\[\equiv (@\sigma_l +0) f (@\sigma_r +0) \mbox{ This is the generic form to build.} \]

Bracket Notation relates to the ONF, or how to build the code, through $\gamma$ and pseudo-code, i.e., 
$\mbox{rav} (\vec i \psi \xi) \equiv (\mbox{rav} \;\xi)[\gamma (\vec i; \rho \xi)]$. With a family of gamma, $\gamma$, functions, e.g., row major, column major, a Cartesian index is related to an offset of the array in memory laid out contiguously. {\bf rav} flattens an array based on it's layout. 
Subscripts relate to left or right arguments and superscripts specify dimensionality. 

\vspace{.2cm}
\noindent
With this, introduce the important identify that is true in general for n-d arrays, denoted by $\xi$:
\[( \iota (\rho \xi)) \psi \xi ) \equiv \xi \] 

{\it This means, with an array's shape, $\rho \xi$, generate an array of indices, $\iota (\rho \xi)$. Then, using that array as an argument to Psi, the original array $\xi$, is returned.}

\vspace{.2cm}
\noindent
In the scalar case, where $\sigma$ denotes a scalar, 
\[ ((\iota (\rho \sigma)) \psi \sigma ) \equiv \sigma \] 
The shape of a scalar is the empty vector $<>$
and, and the only valid index a scalar has is the empty vector, $<>$.
\[ (\iota <>) \psi \sigma \equiv \sigma \]

\[ <> \psi \sigma \equiv \sigma \]
Thus, as long as we can get shapes, e.g., $\rho \xi$, we can recover
all indices from shapes, e.g., $\iota (\rho \xi))$
that have the properties of the $\psi$ function. This is true in general, no mater what the dimensionality of the array including scalars, a 0-dimensional array\cite{mul88}.

Building upon scalar operations, and extending to scalar vector operations, introduced as {\em scalar extension}, later generalized  as the {\em outer product}. The generalization is completed by adding  {\em reductions/contractions} and, the {\em inner product} (defined using reduction and outer product). 
We connect these four ideas formulated mathematically in MoA; scalar operations: Matrix Multiplication (MM), Hadamard Product (HP), and the Kronecker Product (KP) using {\bf one} algorithm/circuit (ipophp) \cite{Grout18,Grout19}. These designs were extended to {\em Blocked} matrix matrix multiplication by first formulating in MoA, then derived, built, and  tested in C, proving it's performance exceeds modern DGEMM  libraries\cite{Mm1,Mm2}. From these experiments, it became clear that a general purpose machine, e.g., cache memory design, languages, Operating Systems (OSes), compilers, ..., would not suffice for optimal array computation. Memory  management, resource management, and the ability to control and optimize array computation was difficult if not impossible.
Much information about the machines was either inaccurate, or was not divulged. Consequently, guided experiments with sophisticated scripts that overlay performance plots, helps  to obtain essential, unknown variables in a developing theoretical MoA model of computation. This information guides the research herein. Research continues to validate that the more shape aware the Operational Normal Form (ONF) is of data and hardware, the easier it is to use special purpose hardware to obtain large amounts/blocks of data(strided DMAs and RMAs, buffers, PIMs,...), that could easily  use  pre-scheduled information of sizes and speeds, up and down the memory hierarchy, {\em deterministically}.

MoA defines all operations using shapes and the Psi indexing function. Hence, when operations are formulated in the MoA algebra, it can be reduced through {\em Psi Reduction} to a semantic/denotational normal form (DNF), i.e., Cartesian indices of all operations are composed.
Then, one of the mapping functions, e.g., $\gamma_{row}$\footnote{There are a whole family of $\gamma$, layout functions: $\gamma_{row},\gamma_{col}, \gamma_{sparse}, etc.$}, transforms an index to an offset in memory, the DNF, or how to build it with knowledge of data layout that capitalizes on start-stops-stride values. \cite{psi-corr}.
Next, formulating dimension lifting, i.e., partitioning shapes, leads to mapping data to hardware. 
 Combined with Lambda Calculus, iteration, sequence, and control \cite{Berkling90}, we have a Turing complete paradigm to reason about array computation in general.  These fundamentals allow one to define and optimize programs in any domain that use the algebra
 and could be enhanced subsets of 
any language with arrays, e.g., Fortran, Python, or Julia \cite{pythonmoa}. 
 That is, when there is {\bf semantic equivalence} across  programming languages, soft or hard, automatic linear and multi-linear transformations are easily applied, proving correctness by construction. MoA is a {\em Universal Algebra} than can optimize all domain specific languages that use arrays/tensors as their primary data structure.

\subsection*{Why MoA Inner and Outer Products?}
Tensor contractions are 3-d extensions of the matrix multiplication \cite{contraction}. Thus, optimizations for higher dimensional
contractions using plus and times (addition and multiplication) are also needed.
In MoA, the formulation of  the inner uses the outer product, noting the  degenerate form of the outer product, is scalar operations. When this is combined with reduction/contraction, i.e., reducing a dimension through addition, or other scalar operation, we have higher dimensional HP, KP, and MMs in one algorithm/circuit (ipophp) or many parallel 2-d versions.


\end{document}